\documentclass[amsmath,10pt,twocolumn,superscriptaddress,footinbib,pre]{revtex4-2}

\pdfoutput=1

\usepackage{amsmath,amsfonts,bm}
\usepackage{graphicx}
\usepackage{amssymb} 
\usepackage{halloweenmath} 
\usepackage{uri}

\usepackage[dvipsnames,usenames]{xcolor}

\usepackage{color}
\usepackage[colorlinks=true,allcolors=blue]{hyperref}
\usepackage{tikz}
\usepackage[normalem]{ulem}

\newcommand{\corr}[1]{\textcolor{black}{#1}}
 
\begin{document}
\title{Thermodynamic uncertainty relation for Langevin dynamics by scaling time}
\author{Rueih-Sheng Fu}
\author{Todd R.~Gingrich}
\email{todd.gingrich@northwestern.edu}
\affiliation{Department of Chemistry, Northwestern University, 2145 Sheridan Road, Evanston, Illinois 60208, USA}

\begin{abstract}
  The thermodynamic uncertainty relation (TUR) quantifies a relationship between current fluctuations and dissipation in out-of-equilibrium overdamped Langevin dynamics, making it a natural counterpart of the fluctuation-dissipation theorem in equilibrium statistical mechanics.
  For underdamped Langevin dynamics, the situation is known to be more complicated, with dynamical activity also playing a role in limiting the magnitude of current fluctuations.
  Progress on those underdamped TUR-like bounds has largely come from applications of the information-theoretic Cram\'er-Rao inequality.
  Here, we present an alternative perspective by employing large deviation theory.
  The approach offers a general, unified treatment of TUR-like bounds for both overdamped and underdamped Langevin dynamics built upon current fluctuations achieved by scaling time.
  The bounds we derive following this approach are similar to known results but with differences we discuss and rationalize. 
\end{abstract}
\maketitle

\section{INTRODUCTION}
Nonequilibrium systems are characterized by currents and dissipation.
In small systems, fluctuations in both quantities can be significant.
Over the last two decades, the mathematical framework of large deviation theory has brought new understanding about the restrictions imposed on these fluctuations.
One notable advance was the development of fluctuation theorems, which express a symmetry relating positive fluctuations to their negative counterparts \cite{kurchan1998fluctuation, crooks1999entropy, jarzynski2000hamiltonian, evans2002fluctuation, seifert2005entropy, andrieux2007fluctuation}.
This symmetry is particularly notable because it holds even outside the linear-response regime.
More recently, large-deviation techniques have been used to derive the thermodynamic uncertainty relation (TUR), an inequality revealing that steady-state fluctuations of a current \(j\) cannot be arbitrarily small.
Rather, these fluctuations are restricted by the system's accumulated entropy production \(\Sigma\) according to the bound
\begin{equation}
	\label{eq:TUR}
	\frac{\text{var}(j)}{\langle j\rangle^2} \geq \frac{2}{\langle\Sigma\rangle}.
\end{equation}

First derived for a Markov jump process in the long-time limit~\cite{barato2015thermodynamic, gingrich2016dissipation, pietzonka2016universal}, the TUR has since been extended to a family of results for finite-time systems~\cite{horowitz2017proof, pietzonka2017finite}, Markov chains~\cite{shiraishi2017finite, proesmans2017discrete}, diffusions~\cite{polettini2016tightening, gingrich2017inferring, hyeon2017physical, van2019uncertainty, lee2021universal, kwon2021thermodynamic, fischer2018large}, periodic driving~\cite{koyuk2018generalization}, and quantum systems~\cite{agarwalla2018assessing, carollo2019unraveling, guarnieri2019thermodynamics, hasegawa2020quantum, hasegawa2021thermodynamic}, along with further generalizations~\cite{pigolotti2017generic, dechant2018multidimensional, niggemann2020field} and specializations~\cite{vroylandt2020isometric, dechant2022bounds}.
We now understand that, in its most general form, a TUR holds because of an involutive symmetry in the system's equation of motion, the most prominent of which is time reversal~\cite{hasegawa2019fluctuation, timpanaro2019thermodynamic, falasco2020unifying}. The involutive symmetry alone suffices to bound current fluctuations as
\begin{equation}
	\label{eq:genTUR}
	\frac{\text{var}(j)}{\langle j\rangle^2} \geq \frac{2}{e^{\langle\Sigma\rangle}-1},
\end{equation}
which is equivalent to Eq.~\eqref{eq:TUR} in the short-time limit but provides essentially no information in the long-time limit~\cite{horowitz2020thermodynamic}. Arriving at a TUR like Eq.~\eqref{eq:TUR} that governs the long-time limit requires that the symmetry be supplemented with additional structure from the equation of motion.
For Markov jump processes and overdamped Langevin dynamics, this additional structure arises from the ability to compute the likelihood of realizing current fluctuations in a particular manner by collectively scaling all microscopic currents.
Since current fluctuations can be realized in that manner or in many other ways, the fluctuations have to be at least as large as the current-scaling construction prescribed.

The corresponding result for underdamped Langevin dynamics is significantly harder to derive.
Simple extensions of the overdamped derivation do not work because the probability of scaling all microscopic currents cannot be computed in the underdamped regime.
Significant progress on this front has come from constructing a virtual perturbation of the dynamics in terms of some parameter \(\theta\), then applying the Cram\'er-Rao bound on said parameter to establish an inequality~\cite{dechant2018multidimensional, hasegawa2019uncertainty, van2019uncertainty, guarnieri2019thermodynamics, lee2021universal, kwon2021thermodynamic, dechant2022bounds}.
In this paper, we derive a similar underdamped inequality from a large-deviation perspective, akin to \cite{fischer2018large}.
Our derivation is based on constructing potentially suboptimal ways to realize current fluctuations by scaling the equation of motion in time, a procedure that can be applied in a physically transparent manner to both the overdamped and underdamped settings.
This notion of scaling time has been fruitful in deriving bounds \cite{garrahan2017simple,fischer2018large,whitelam2020evolutionary} and in numerical sampling \cite{jacobson2019direct}.
In contrast to prior work on a one-dimensional ring~\cite{fischer2018large}, we have pursued this approach at the level of trajectories, which gives rise to more general bounds.

\section{Results}
\subsection{TUR from contraction}
\label{sec:Contraction}
From the large deviation perspective, the TUR is fundamentally built around the contraction principle.
In many cases, the probability of measuring a current \(j\) in an observation time \(\tau\) adopts the long-time asymptotic form
\begin{equation}
    \label{eq:asympprob}
  \rho(j) \asymp e^{-\tau I(j)},
\end{equation}
where \(I(j)\) is the large-deviation rate function for the current fluctuations and the sub-exponential contribution from the prefactor can be neglected in the long-time limit.
This rate function has a minimum at \(\langle j\rangle\), and its value at any given \(j\) reflects how unlikely it is to realize a trajectory with that value of the current on an exponential scale.
For all but the simplest systems, the explicit derivation of \(I(j)\) is intractable and numerical computation is challenging.
This difficulty arises because, in the long-time limit, \(I(j)\) depends only on the probability of the most likely realization yielding a current \(j\);
solving for that optimal realization is generally prohibitively difficult.
On the other hand, it is easy to construct a subset of all possible realizations, and an upper bound on \(I(j)\) follows if one can compute the likelihood of sampling those particular realizations.

Such a bound was developed for Markov jump processes~\cite{gingrich2016dissipation} and extended to overdamped Langevin dynamics~\cite{gingrich2017inferring}.
This was accomplished by scaling the microscopic steady-state currents by a factor \(\eta\) to generate a realization of the system with macroscopic current \(\langle j\rangle/\eta\).
This construction holds the system's empirical density fixed and varies its empirical current as a function of the scale parameter \(\eta\), from which we ultimately obtain the bound
\begin{equation}
    I(j) \leq I_{\eta}(j) \equiv \left(\frac{1}{\eta}-1\right)^2 \frac{\langle\Sigma\rangle}{4\tau},
\end{equation}
where \(\eta = \langle j\rangle/j\) and \(\tau\) is the observation time. The TUR, Eq.~\eqref{eq:TUR}, then follows from the identity \(\text{var}(j) = 1 / I''(\langle j\rangle)\) \cite{touchette2009large}.

This analysis worked for Markov jump processes and overdamped Langevin dynamics because, exceptionally, the large-deviation rate functions for the empirical density and current (level 2.5) were known explicitly for those systems.
In contrast, the level-2.5 rate function is not known for underdamped Langevin dynamics, and the underdamped TUR cannot be derived following the same approach. 
We suggest that this difficulty can be overcome by working with trajectories, in which case an expression for the level-3 underdamped rate functional is known.
Motivated by the appearance of \(\eta\) as a scale parameter for the current in the overdamped regime, we re-interpret \(\eta\) in this new context as a scale parameter for \emph{time}.
In other words, we generate scaled trajectories with a timestep of \(\eta\Delta t\) rather than a timestep of \(\Delta t\). 
As before, this construction holds the system's density fixed and varies its current, again with \(j = \langle j\rangle/\eta\).

The central aim of this paper is to construct large-deviation bounds---and the corresponding uncertainty relations---by computing the asymptotic probability of current fluctuations realized via scaling time in this fashion.
This time-scaling procedure offers a way to extend the large-deviation perspective from overdamped to underdamped dynamics.

\subsection{Overdamped Langevin dynamics}

We analyze a \(d\)-dimensional system, working in discrete time and ultimately recovering continuous-time results by taking the limit \(\Delta t \to 0\). In this limit, the collection of discrete points \(\{\mathbf{x}_i\}\) becomes the continuous trajectory \(\mathbf{x}(t)\). Similarly, the collection of noise \(\{\boldsymbol{\xi}_i\}\) becomes the continuous-time white-noise process \(\boldsymbol{\xi}(t)\).
The overdamped Langevin equation is given by 
\begin{equation}
    \gamma\dot{\mathbf{x}} = \mathbf{F}(\mathbf{x}) + \sqrt{2\gamma T}\boldsymbol{\xi},
\end{equation}
or in discretized form as
\begin{equation}
	\label{eq:contoverdampedLangevin}
	\gamma\Delta \mathbf{x}_i = \mathbf{F}_i\Delta t + \sqrt{2\gamma T\Delta t}\boldsymbol{\xi}_i,
\end{equation}
where the subscript \(i\) indexes timesteps, \(\gamma\) is the friction coefficient, \(\mathbf{F}_i \equiv \mathbf{F}(\mathbf{x}_i)\) is a position-dependent force, and \(T\) is the temperature. The standard Gaussian noise \(\boldsymbol{\xi}_i\) satisfies \(\langle\boldsymbol{\xi}_i\rangle = 0\) and \(\langle\boldsymbol{\xi}_i\boldsymbol{\xi}_j\rangle = \delta_{ij}\mathbf{I}_d\), where \(\mathbf{I}_d\) is the \(d\)-dimensional identity matrix.
Solutions of this equation over a time interval \(\tau = N\Delta t\) are stochastic trajectories \(\{\mathbf{x}_i\}\) parametrized by \(i\).

Using the large-deviation machinery explained in the last section, we will derive our first main result,
\begin{equation}
	\label{eq:overdampedmodTUR}
	\frac{\text{var}(j)}{(\langle j\rangle - \kappa T)^2} \geq \frac{2\gamma T}{\tau\langle\mathbf{F}^2\rangle} = \frac{2}{\langle\Sigma\rangle - \tau\langle\nabla\cdot\mathbf{F}\rangle/\gamma},
\end{equation}
where \(\kappa = \text{d}\langle j\rangle/\text{d}T\). 
Unless otherwise stated, all continuous-time expectation values are to be interpreted following the It\^{o} convention.
For instance, the average squared force \(\langle\mathbf{F}^2\rangle\) is given by
\begin{equation}
	\langle \mathbf{F}^2\rangle = \frac{1}{N}\sum_{i=0}^{N-1}\mathbf{F}_i^2.
\end{equation}

Both the left-hand and right-hand sides of Eq.~\eqref{eq:overdampedmodTUR} differ slightly from the standard TUR for overdamped Langevin processes, Eq.~\eqref{eq:TUR}.
The right-hand side includes an additive term \(-\tau\langle\nabla\cdot\mathbf{F}\rangle/\gamma\) in the denominator. If \(\langle\nabla\cdot\mathbf{F}\rangle = 0\), this term vanishes and we recover the right-hand side of Eq.~\eqref{eq:TUR}.
One way that this expectation value can vanish is if the divergence vanishes throughout space, as in a solenoidal vector field, one in which there are no sources or sinks.
Eq.~\eqref{eq:TUR} also differs by the inclusion of the \(-\kappa T\) term on the left-hand side of the bound. Because the current typically increases with temperature, this term generally weakens our bound relative to the usual TUR, though our bound is strengthened in the unusual situation in which \(\kappa < 0\).

\begin{figure}
	\includegraphics[width=0.4\textwidth]{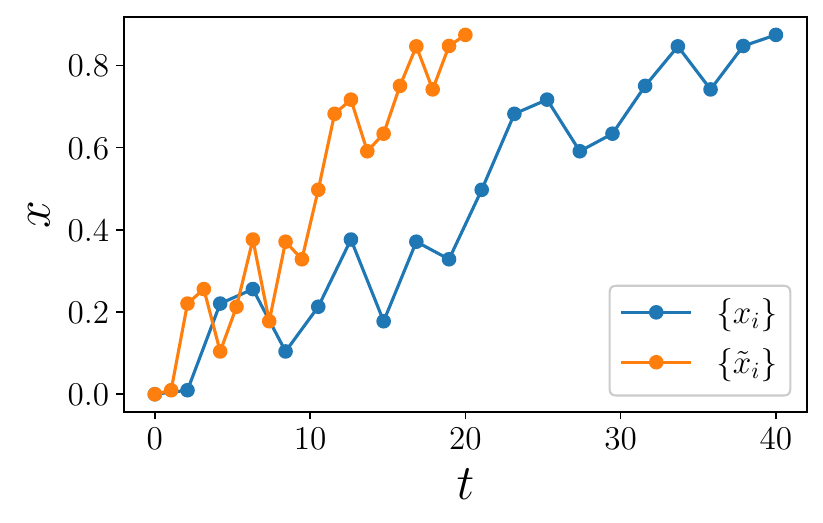}
	\centering
	\caption{Schematic of time-scaling construction. The original trajectory in blue, with timestep \(\Delta t = 1\) is scaled by a factor \(\eta = 0.5\) to generate a scaled trajectory in orange, which has timestep \(\eta\Delta t = 0.5\). Both trajectories visit the same points in space, but the orange trajectory is sped up by a factor of two, and thus has twice the current.}
	\label{fig:scaling}
\end{figure}

To derive the first main result, we consider a trajectory \(\{\mathbf{x}_i\}\) and construct a corresponding scaled trajectory \(\{\tilde{\mathbf{x}}_i\}\) that visits the same discrete points with a scaled timestep (see Fig.~\ref{fig:scaling}).
That scaled trajectory satisfies the scaled equation of motion (with timestep \(\eta\Delta t\))
\begin{equation}
	\gamma\Delta\tilde{\mathbf{x}}_i = \mathbf{F}_i\eta\Delta t + \sqrt{2\gamma T\eta\Delta t}\tilde{\boldsymbol{\xi}}_i,
\end{equation}
where \(\tilde{\boldsymbol{\xi}}_i\) samples the same unit normal distribution as \(\boldsymbol{\xi}_i\).
Enforcing the constraint that the two trajectories visit identical points in phase space (\(\Delta\mathbf{x}_i = \Delta\tilde{\mathbf{x}}_i\)), though with different clocks, links the noises \(\boldsymbol{\xi}_i\) and \(\tilde{\boldsymbol{\xi}}_i\) via their respective equations of motion to yield
\begin{equation}
  \boldsymbol{\xi}_i = \sqrt{\eta}\tilde{\boldsymbol{\xi}}_i + \mathbf{F}_i\sqrt{\frac{\Delta t}{2\gamma T}}(\eta - 1).
\end{equation}

By itself, this transformation is not very useful because of the scale factor \(\sqrt{\eta}\) in front of \(\tilde{\boldsymbol{\xi}}_i\). As a result, the trajectories \(\{\mathbf{x}_i\}\) and \(\{\tilde{\mathbf{x}}_i\}\) live in different spaces and are not directly comparable. More technically, reweighting the trajectories is not possible because \(\{\boldsymbol{\xi}_i\}\) and \(\{\tilde{\boldsymbol{\xi}}_i\}\) have different diffusion constants and are thus not mutually absolutely continuous. To resolve this issue, we simultaneously scale the temperature as \(T \to T/\eta\), obtaining the revised equivalence
\begin{equation}
    \label{eq:scalednoise}
	\boldsymbol{\xi}_i = \tilde{\boldsymbol{\xi}}_i + \mathbf{F}_i\sqrt{\frac{\Delta t}{2\gamma T}}(\eta - 1).
\end{equation}
This time and temperature scaling suggests that we consider the parametric rate function \(I(j; T)\) in which we highlight the role of \(T\) as an argument explicitly. In Appendix~\ref{app:twolemmas}, we argue that this rate function can be bounded in the form
\begin{equation}
    \label{eq:scaledoverdampedratefunctional}
	I\left(\frac{\langle j\rangle_T}{\eta};\frac{T}{\eta}\right) \leq \corr{-}\frac{1}{\corr{2}\eta\tau}\sum_{i=0}^{N-1}\langle\tilde{\boldsymbol{\xi}}_i^2 - \boldsymbol{\xi}_i^2\rangle = \frac{(\eta - 1)^2}{\eta}\frac{\langle \mathbf{F}^2\rangle}{4\gamma T}.
\end{equation}

\begin{figure}
	\includegraphics[width=0.4\textwidth]{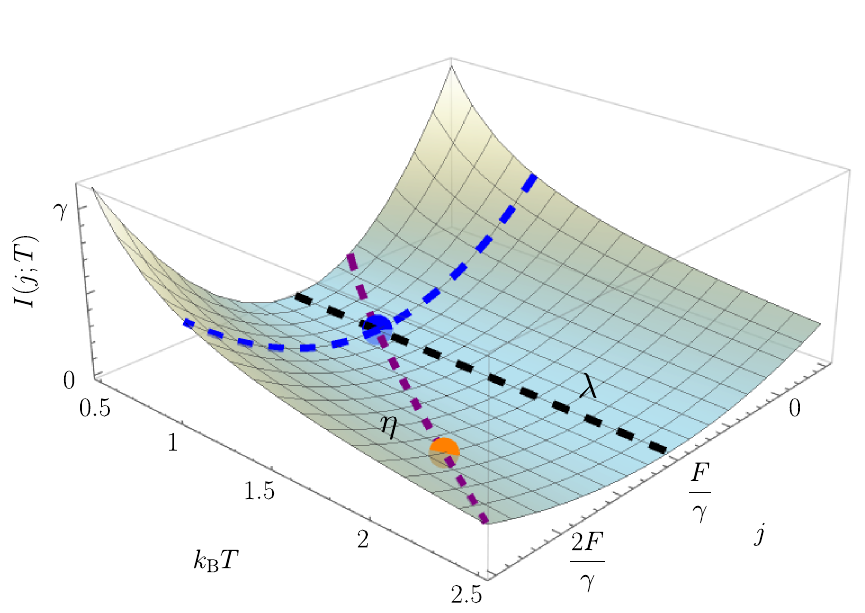}
	\centering
	\caption{Schematic of a coordinate transformation from \((T, j)\) to \((\lambda, \eta)\), with \(\lambda\) and \(\eta\) as defined in the main text. We exemplify the transformation with the rate function for a one-dimensional constant-force system with \(F(x) \equiv F\). In this case, the average current is \(\langle j\rangle = F/\gamma\), independent of the temperature \(T\), and the rate function \(I(j; T)\) is given by \(I(j; T) = (\gamma/4T)(j-F/\gamma)^2\). The induced local coordinate transformation replaces the orthogonal \(T\)- and \(j\)-axes (black and blue dashed lines) with the non-orthogonal \(\lambda\)- and \(\eta\)-axes (black and purple dashed lines). The value of the rate function at the blue and orange points respectively includes contributions from the blue and orange trajectories of Fig.~\ref{fig:scaling}.}
	\label{fig:coordtrans}
\end{figure}

Ultimately, we would like to---and we shall see that we can---use the bound Eq.~\eqref{eq:scaledoverdampedratefunctional} to derive a corresponding bound on \(I(j; T)\). To do so, we pick a reference temperature \(T_0\) and perform a local transformation about the point \((T_0, \langle j\rangle_{T_0})\) from the variables \(j\) and \(T\) to \(\eta\) and \(\lambda\), coordinates that are more natural to the system. \(\eta\) and \(\lambda\) parametrize two lines in the \((T, j)\)-plane.
See Fig.~\ref{fig:coordtrans}.
The line parametrized by \(\eta\) represents a simultaneous scaling of time and temperature along which the bound Eq.~\eqref{eq:scaledoverdampedratefunctional} is known. It passes through \((T_0, \langle j\rangle_{T_0})\) and the origin and takes the form \((T_0/\eta, \langle j\rangle_{T_0}/\eta)\).
The line parametrized by \(\lambda\) is the tangent line to the curve \(\langle j\rangle(T) \equiv \langle j\rangle_T\), the average current as a function of the temperature. Along this curve, the rate function and therefore all its derivatives vanish. Except when the current is a linear function of the temperature, this choice of \(\lambda\) provides a second linearly independent variable along which we can bound \(I(j; T)\).
We parametrize the curve \(\langle j\rangle_T\) as \((\lambda T_0, \langle j\rangle_{\lambda T_0})\), and hence its tangent line as \((\lambda T_0, \langle j\rangle_{T_0} + \kappa(\lambda - 1)T_0)\), where \[\kappa := \left.\frac{\text{d}\langle j\rangle_{\lambda T_0}}{\text{d}T}\right|_{\lambda = 1}.\]

In Appendix~\ref{app:twolemmas}, we show that the fact that \(I\) and its derivatives vanish along the curve \(\langle j\rangle_T\) imply that \(I\) and its first two derivatives with respect to the tangent line vanish at \((T_0, \langle j\rangle_{T_0})\). This is important because it means we have access to information about the derivatives of \(I\) along \(\lambda\) and \(\eta\). As long as these two lines are independent---as long as \(\langle j\rangle_T\) is not linear in \(T\)---we can also obtain information about the derivatives of \(I\) along \(j\) by linear transformation. 

By taking appropriate derivatives of the parametrization of the two lines corresponding to \(\lambda\) and \(\eta\), we find that 
\begin{equation}
	\begin{bmatrix} \text{d}T\\ \text{d}j \end{bmatrix} = \begin{bmatrix} T_0 & -T_0\\ \kappa T_0 & -\langle j\rangle_{T_0} \end{bmatrix} \begin{bmatrix} \text{d}\lambda\\ \text{d}\eta \end{bmatrix}.
\end{equation}
Matrix inversion yields
\begin{equation}
	\begin{bmatrix} \text{d}\lambda\\ \text{d}\eta \end{bmatrix} = \frac{1}{\kappa T_0^2 - \langle j\rangle_{T_0}T_0}\begin{bmatrix} -\langle j\rangle_{T_0} & T_0\\ -\kappa T_0 & T_0 \end{bmatrix} \begin{bmatrix} \text{d}T\\ \text{d}j \end{bmatrix}.
\end{equation}
It remains to perform a Taylor expansion of \(I\) about \((T_0, \langle j\rangle_{T_0})\). To second order, we find
\begin{align}
	I(j; T) &= \frac{(j - \langle j\rangle_{T_0})^2}{2}\left.\frac{\partial^2 I}{\partial j^2}\right|_{(T_0, \langle j\rangle_{T_0})}\\ \nonumber
	    &\qquad \qquad+ \frac{(T - T_0)^2}{2}\left.\frac{\partial^2 I}{\partial T^2}\right|_{(T_0, \langle j\rangle_{T_0})}.
\end{align}
Expanding the derivatives in the \((\lambda, \eta)\) basis and recognizing that partial derivatives with respect to \(\lambda\) vanish, we have
\begin{align}
	\frac{\partial^2}{\partial j^2} &= \left(\frac{1}{\kappa T_0-\langle j\rangle_{T_0}}\right)^2\frac{\partial^2}{\partial\eta^2},\\
	\frac{\partial^2}{\partial T^2} &= \left(\frac{-\kappa}{\kappa T_0-\langle j\rangle_{T_0}}\right)^2\frac{\partial^2}{\partial\eta^2},
\end{align}
and hence see that
\begin{align}
	I(j; T) &= \frac{\kappa^2(T-T_0)^2 + (j-\langle j\rangle_{T_0})^2}{2(\kappa T_0 - \langle j\rangle_{T_0})^2}\left.\frac{\partial^2 I}{\partial\eta^2}\right|_{\eta = 1}\\
        &\leq \frac{(j - \langle j\rangle)^2}{(\langle j\rangle - \kappa T)^2}\frac{\langle \mathbf{F}^2\rangle}{4\gamma T} \label{eq:newscaledoverdampedratefunctional}
\end{align}
upon using Eq.~\eqref{eq:scaledoverdampedratefunctional}, specializing to \(T = T_0\), and taking \(\langle j\rangle \equiv \langle j\rangle_T\). 
Expanding \(I(j; T)\) about \(\langle j\rangle\) to leading order then yields
\begin{equation}
    \label{eq:rearrangedoverdampedbound}
	\frac{1}{\text{var}(j)} = I''(\langle j\rangle) \leq \frac{1}{(\langle j\rangle - \kappa T)^2}\frac{\langle\mathbf{F}^2\rangle}{2\gamma T},
\end{equation}
noting that \(I(\langle j\rangle) = I'(\langle j\rangle) = 0\).
Rearranging Eq.~\eqref{eq:rearrangedoverdampedbound} gives our first main result, Eq.~\eqref{eq:overdampedmodTUR}.
The inequality can be re-expressed in terms of the entropy production by recognizing that the entropy production \(\langle\Sigma\rangle\) is the time-antisymmetric part of the action, \(\tau I(\{\mathbf{x}_i\})\), and takes the form
\begin{align}
	\langle\Sigma\rangle &= \frac{1}{2T}\sum_i\langle(\mathbf{F}_i + \mathbf{F}_{i+1})\cdot\Delta\mathbf{x}_i\rangle\\
	    &\approx \frac{1}{T}\sum_i\langle\mathbf{F}_i\cdot\Delta\mathbf{x}_i\rangle + \frac{1}{2T}\sum_i\langle\Delta\mathbf{x}_i\cdot\nabla\mathbf{F}_i\cdot\Delta\mathbf{x}_i\rangle\\
	    &= \frac{\tau\langle\mathbf{F}^2\rangle}{\gamma T} + \frac{\tau}{\gamma}\langle\nabla\cdot\mathbf{F}\rangle,
\end{align}
where the first-order Taylor expansion of \(\mathbf{F}_{i+1}\) about \(\mathbf{x}_i\) becomes exact as \(\Delta t \to 0\).

Our overdamped bound, Eq.~\eqref{eq:overdampedmodTUR}, reproduces prior work~\cite{gingrich2016dissipation} when both \(\langle\nabla\cdot\mathbf{F}\rangle\) and \(\kappa\) vanish, but the negativity of \(\langle\nabla\cdot\mathbf{F}\rangle\), derived in Appendix~\ref{app:divF}, generically weakens the bound.
It is natural to consider why the time-scaling bound would be weaker than the current-scaling TUR.
Both large-deviation bounds require that we pass from a high-dimensional distribution to the single-variable distribution over a current.
In our time-scaling construction, the high-dimensional level-3 distribution is over trajectories.
By contrast, the current-scaling construction involves a level-2.5 distribution over densities and currents; each realization of the system at this level of description corresponds to multiple trajectories.
By scaling the currents at level 2.5, rather than the trajectories themselves at level 3, we can consider a larger subset of all possible trajectories and hence generate a tighter bound.
Notably, in the constant-force scenario of Fig.~\ref{fig:coordtrans}, \(\kappa = \langle \nabla \cdot \mathbf{F}\rangle = 0\), the same bound is obtained from both the level-3 and level-2.5 descriptions.

\subsection{Underdamped Langevin dynamics}
We repeat the analysis for underdamped Langevin dynamics,
\begin{equation}
    m\ddot{\mathbf{x}} = -\gamma\dot{\mathbf{x}} + \mathbf{F}(\mathbf{x}) + \sqrt{2\gamma T}\boldsymbol{\xi},  
\end{equation}
with unit mass.
Following the discretization scheme in \cite{gronbech2013simple}, we have
\begin{align}
	\label{eq:contunderdampedLangevin}
	\Delta \mathbf{x}_i &= b\Delta t\left(\dot{\mathbf{x}}_i + \frac{\mathbf{F}_i\Delta t}{2}\right) + \frac{b\Delta t}{2}\sqrt{2\gamma T\Delta t}\boldsymbol{\xi}_i,\\
	\Delta \dot{\mathbf{x}}_i &= \frac{\Delta t}{2}(\mathbf{F}_i + \mathbf{F}_{i+1}) - \gamma\Delta \mathbf{x}_i + \sqrt{2\gamma T\Delta t}\boldsymbol{\xi}_i, \label{eq:contunderdampedLangevintwo}
\end{align}
where \(b \equiv (1+\gamma\Delta t/2)^{-1}\). As before, we generate a scaled trajectory \(\{\tilde{\mathbf{x}}_i\}\) which is spatially identical to the unscaled trajectory, with \(\tilde{\mathbf{x}}_i = \mathbf{x}_i\). In the underdamped regime, scaling time requires that we scale the velocity in the same fashion, generating the additional constraint \(\dot{\mathbf{x}}_i = \eta\dot{\tilde{\mathbf{x}}}_i\). The same procedure as in the overdamped regime yields our second main result, 
\begin{align}
	\label{eq:underdampedmodTURprecursor}
	\frac{\text{var}(j)}{(\langle j\rangle - \corr{3}\kappa T)^2} &\geq \frac{2\gamma T}{\tau(4\langle \mathbf{F}^2\rangle - 3\gamma^2\langle\dot{\mathbf{x}}^2\rangle + 4\gamma^2T)}\\
	    &= \frac{2}{16\langle\Upsilon\rangle + 9\langle\Sigma\rangle - 3\gamma\tau},\label{eq:underdampedmodTUR}
\end{align}
where \(\langle\Upsilon\rangle\) is the dynamical activity, the time-symmetric part of the action that excludes the functional measure~\cite{falasco2016nonequilibrium}. See Appendix~\ref{app:actionquantities} for more details. Our bound mirrors that of \cite{fischer2018large, van2019uncertainty}, but with an additional term \(-3\gamma\tau\) in the denominator on the right and an additional term \(-\corr{3}\kappa T\) in the denominator on the left. The former term tightens our bound, particularly for large \(\gamma\), whereas the latter generally weakens our bound. The two effects are independent and are of varying strength. 

The dynamical activity does not appear in the overdamped bound because, in that regime, the entropy production and dynamical activity are Legendre duals and hence not independent~\cite{maes2008steady, maes2017frenetic}. More fundamentally, the probability currents in overdamped Langevin dynamics can be described solely in terms of variables invariant under time reversal, corresponding to the irreversible currents that suffice to characterize the entropy production. In contrast, underdamped Langevin dynamics also includes variables antisymmetric under time reversal, corresponding to reversible currents that characterize the dynamical activity. This distinction explains the appearance of the dynamical activity in the underdamped but not the overdamped bound \cite{spinney2012entropy}. 

By analogy with Eq.~\eqref{eq:scalednoise}, we identify a relationship between the scaled and unscaled noises by applying the constraints \(\Delta\mathbf{x} = \Delta\tilde{\mathbf{x}}\) and \(\Delta\dot{\mathbf{x}} = \eta\Delta\dot{\tilde{\mathbf{x}}}\). \corr{In the underdamped case, because the stochastic part of the displacement scales as \(\sqrt{T(\Delta t)^3}\boldsymbol{\xi}_i\) (cf.~Eq.~\eqref{eq:contunderdampedLangevin}), we must scale the temperature as \(T \to T/\eta^3\). The resulting} relationship is given by 
\begin{align}
    \label{eq:underdampedscalednoise}
	\boldsymbol{\xi}_i &= \corr{\epsilon}\tilde{\boldsymbol{\xi}}_i \corr{+} \frac{2\dot{\mathbf{x}}_i}{\sqrt{2\gamma T\Delta t}}\corr{(\epsilon-1) +} \mathbf{F}_i\sqrt{\frac{\Delta t}{2\gamma T}}\corr{(\epsilon\eta^2-1)}\\
	    &= \corr{\tilde{\boldsymbol{\xi}}_i + \sqrt{\frac{\Delta t}{2\gamma T}}(\mathbf{F}_i(\eta^2-1) - \gamma\dot{\mathbf{x}}_i(\eta-1)) + O(\Delta t)},
\end{align}
where
\begin{equation}
	\epsilon = \frac{1+\gamma\Delta t/2}{1+\gamma\eta\Delta t/2} = \frac{1}{1+(1-b)(\eta-1)}\corr{.}
\end{equation}
\corr{Keeping only leading-order terms in \(\Delta t\), we find}
\begin{align}
	I\left(\frac{\langle j\rangle_T}{\eta};\frac{T}{\eta\corr{^3}}\right) &\leq \corr{\frac{1}{4\gamma T}\left\langle(\mathbf{F}(\eta^2-1) - \gamma\dot{\mathbf{x}}(\eta-1))^2\right\rangle}\\
	&= (\eta - 1)^2\langle\Xi\rangle + O((\eta - 1)^3) \label{eq:underdampedratefn},
\end{align}
where 
\begin{align}
	\langle\Xi\rangle &= \corr{\frac{1}{4\gamma T}\left\langle(2\mathbf{F}-\gamma\dot{\mathbf{x}})^2\right\rangle}.
\end{align}
Again expanding \(I(j)\) about \(\langle j\rangle\) to leading order and taking the \(\Delta t \to 0\) limit yields
\begin{equation}
	\label{eq:underdampedratefnbound}
	\frac{1}{\text{var}(j)} = I''(\langle j\rangle) \leq \frac{2\langle\Xi\rangle}{(\langle j\rangle - \corr{3}\kappa T)^2}.
\end{equation}
\corr{The coefficient of 3 in Eq.~\eqref{eq:underdampedratefnbound} arises from the modified scaling \(T \to T/\eta^3\).}
Eq.~\eqref{eq:underdampedratefnbound} can be rearranged to give Eq.~\eqref{eq:underdampedmodTURprecursor} by applying the identity
\begin{equation}
    \label{eq:underdampedidentity}
	\langle \mathbf{F}\cdot\dot{\mathbf{x}}\rangle = \gamma\langle\dot{\mathbf{x}}^2\rangle - \gamma T.
\end{equation}
To proceed from Eq.~\eqref{eq:underdampedmodTURprecursor} to Eq.~\eqref{eq:underdampedmodTUR}, we would like to express the bound in terms of the entropy production and dynamical activity. In Appendix~\ref{app:expectationvalues}, we prove Eq.~\eqref{eq:underdampedidentity} and show that for underdamped dynamics,
\begin{align}
	\langle\Sigma\rangle &= \tau\frac{\langle \mathbf{F}\cdot\dot{\mathbf{x}}\rangle}{T} = \tau\frac{\gamma\langle\dot{\mathbf{x}}^2\rangle}{T} - \gamma\tau,\\
	\langle\Upsilon\rangle &= \frac{\tau}{4}\left(\frac{\langle\mathbf{F}^2\rangle}{\gamma T} - \frac{3\gamma\langle\dot{\mathbf{x}}^2\rangle}{T} + 4\gamma\right).
\end{align}
Combining these expressions with Eq.~\eqref{eq:underdampedratefnbound} recovers our second main result, Eq.~\eqref{eq:underdampedmodTUR}.

\subsection{Numerical verification of TUR bounds}

In the overdamped regime, rearranging Eq.~\eqref{eq:newscaledoverdampedratefunctional} gives the nondimensional result
\begin{equation}
	\label{eq:overdampednondbound}
	I^*_\text{over}(j) \equiv \frac{4\gamma T}{\langle\mathbf{F}^2\rangle}\left(1 - \frac{\kappa T}{\langle j\rangle}\right)^2I(j) \leq \left(\frac{1}{\eta} - 1\right)^2.
\end{equation}
Similarly, the underdamped TUR, Eq.~\eqref{eq:underdampedratefnbound}, corresponds to the nondimensional quadratic truncation
\begin{equation}
	\label{eq:underdampednondbound}
	I^*_\text{under}(j) \equiv \left(1 - \frac{\corr{3}\kappa T}{\langle j\rangle}\right)^2\frac{I(j)}{\langle\Xi\rangle} \leq \left(\frac{1}{\eta} - 1\right)^2,
\end{equation}
which must hold in the neighborhood of \(\eta = 1\).

\begin{figure}
	\includegraphics[width=0.45\textwidth]{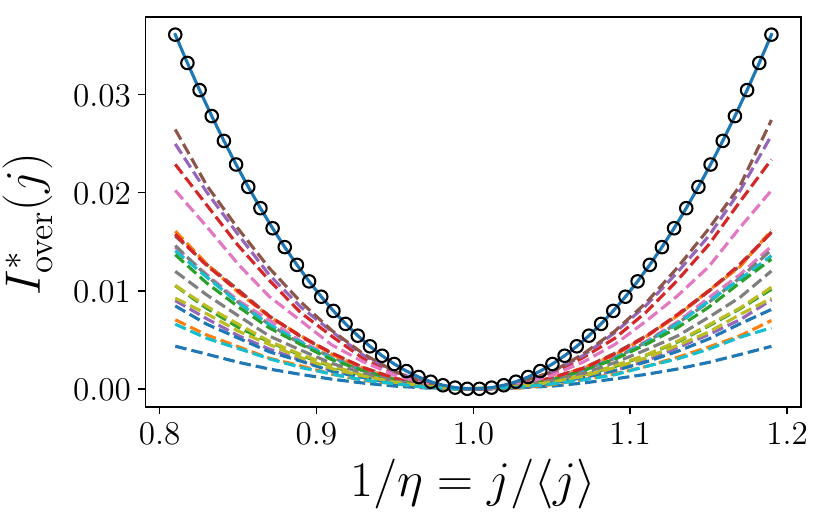}
	\includegraphics[width=0.45\textwidth]{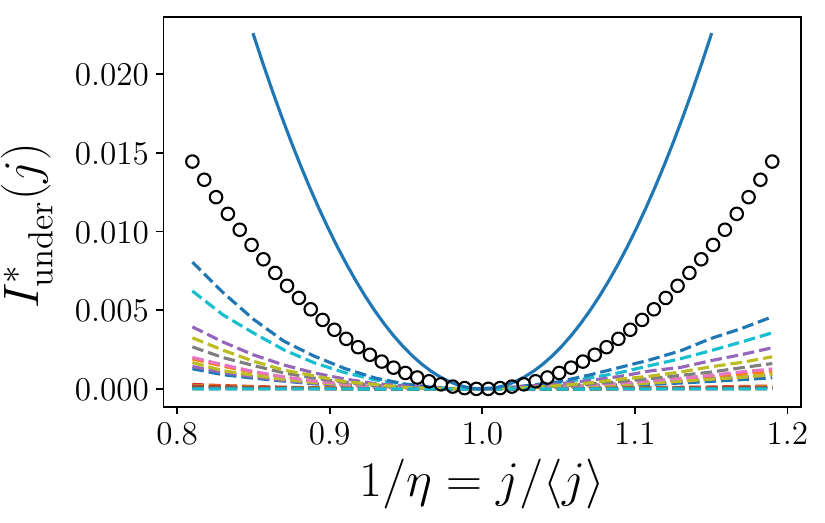}
	\centering
	\caption{(top: overdamped, bottom: underdamped) Nondimensionalized plots of the rate function \(I(j)\) with respect to the scale parameter \(\eta\) for a one-dimensional periodic system subject to the force \(F(x) = 1 + a\sin(x) + b\cos(x) + c\sin(2x) + d\cos(2x)\) for stochastic parameters \(a, b, c, d \in \text{Uniform}[0, 1)\). \(I^*_\text{over}\) and \(I^*_\text{under}\) are nondimensionalized versions of the overdamped and underdamped rate functions defined in Eqs.~\eqref{eq:overdampednondbound} and \eqref{eq:underdampednondbound}. In nondimensional units, the upper bound takes on the form \(y = (x-1)^2\), corresponding to the solid blue curve in both plots.
	The curves composed of black circles represent the constant-force results when \(a = b = c = d = 0\).
    The twenty dashed curves in both plots each reveal fluctuation for a random sampling of parameters \(a\) through \(d\), extracted from 100,000 independent simulations for each choice of parameters with \(N = 1\times 10^7\), \(\Delta t = 0.01\), \(\gamma = 1\), and \(T = 1.5\).
    \(\langle\Xi\rangle\) is given by \(\langle F^2\rangle/(\gamma T) - \langle F\dot{x}\rangle/T + \gamma\langle \dot{x}^2\rangle/(4T)\).
    \(\kappa\) is obtained by linear approximation using the values of \(\langle j\rangle\) at \(T = 1.5\) and \(T = 1.51\).
    These plots verify Eqs.~\eqref{eq:scaledoverdampedratefunctional} and \eqref{eq:underdampedratefn} because, in each case, the solid blue curve lies above all dashed curves in the neighborhood of \(\eta = 1\).}
	\label{fig:plotoverdampednond}
\end{figure}

The time-scaling procedure applies generally to systems with more than one particle in more than one dimension. For simplicity, we verify the bounds Eqs.~\eqref{eq:overdampednondbound} and \eqref{eq:underdampednondbound} numerically for a single particle on a one-dimensional ring.
That particle is subject to a spatially dependent force \(F(x)\) for various choices of \(F(x)\), as illustrated in Fig.~\ref{fig:plotoverdampednond}.
In both plots, the solid blue line representing the bound lies above each of twenty rate functions obtained from numerical simulation. In the overdamped regime, the bound is saturated in the special case of a constant driving force \(F(x) \equiv F\), shown in black circles. In the underdamped regime, the bound is not saturated even in this special case; in this scenario, we have
\begin{equation}
	\corr{I(j) \leq \left(\frac{1}{\eta} - 1\right)^2\left(1 - \frac{3\kappa T}{\langle j\rangle}\right)^{-2}\left(\langle\Xi\rangle - \frac{\gamma}{4}\right),}
\end{equation}
where the additive term \(\gamma/4\) is responsible for the lack of saturation.

We attribute this lack of saturation to the fact that the underdamped equation of motion is of higher order than the overdamped equation of motion. The overdamped equation is first order and stochastic in \(x\), whereas the underdamped equation is second order and stochastic in \(\dot{x}\). This difference in order implies that the derivation of the overdamped Langevin equation from the underdamped one is subtle and cannot be effected by the simple limit \(m \to 0\). Instead, the conventional argument takes the limit \(\gamma \to \infty\) and invokes a separation of timescales between the position and momentum degrees of freedom \cite{pavliotis2014langevin}. The friction coefficient \(\gamma\) is responsible for characterizing one such relevant timescale, so we expect that it will also modulate the bound.

\section{Conclusion}

In this paper, we have described a method of developing TUR-like bounds by introducing a continuous parameter that scales time and generates atypical trajectories which serve to bound the system's rate function. Passing such trajectories through standard large-deviation machinery generates the aforementioned bounds.

This method generalizes that used in the derivation of the TUR. Because it works directly with trajectories, rather than with the reduced densities and currents, it is applicable to both the underdamped and overdamped regimes.
The resulting bounds are comparable but slightly different from known results in the overdamped and underdamped regime, and we rationalize this discrepancy by considering the different levels of description from which these bounds were derived.
We emphasize the general utility of this procedure and suggest that it may be used fruitfully to derive other related bounds.

\section{Acknowledgments}
We gratefully acknowledge Hadrien Vroylandt and Patrick Pietzonka for insightful discussions. An anonymous referee was instrumental in prompting us to develop an approach that scaled temperature along with time so as to resolve a divergence that appeared in our initial draft. Research reported in this publication was supported by the Gordon and Betty Moore Foundation through Grant No. GBMF10790.

\appendix
\section{\(\langle\nabla\cdot\mathbf{F}\rangle < 0\) for overdamped Langevin dynamics} \label{app:divF}
The probability current
\begin{equation}
    \gamma\mathbf{J} = \mathbf{F}\rho - T\nabla\rho
\end{equation}
is a conserved quantity at steady state, since \(\nabla\cdot\mathbf{J} = 0\) by the Fokker-Planck equation. Solving for \(\mathbf{F}\) yields
\begin{equation}
    \mathbf{F} = \frac{\gamma\mathbf{J}}{\rho} + T\nabla\ln\rho,
\end{equation}
and hence
\begin{equation}
    \nabla\cdot\mathbf{F} = -\frac{\gamma\mathbf{J}\cdot\nabla\rho}{\rho^2} + T\nabla^2\ln\rho.
\end{equation}
Taking expectations and integrating by parts, we find
\begin{align}
    \langle\nabla\cdot\mathbf{F}\rangle &= \int\text{d}\mathbf{x}\,\rho(\nabla\cdot\mathbf{F})\\
        &= -\gamma\int\text{d}\mathbf{x}\,\mathbf{J}\cdot\nabla\ln\rho + T\int\text{d}\mathbf{x}\,\rho\nabla^2\ln\rho\\
        &= - T\int\text{d}\mathbf{x}\,\frac{(\nabla\rho)^2}{\rho} < 0.
\end{align}

\section{Two lemmas in the derivation of the time-scaling bound} \label{app:twolemmas}

For simplicity, we limit ourselves to the one-dimensional case, though the generalization to the multidimensional case is straightforward.
\vspace{1ex}

\noindent\textbf{Lemma 1.} A bound on the overdamped rate function is given by
\begin{equation}
	I\left(\frac{\langle j\rangle}{\eta};\frac{T}{\eta}\right) \leq (\eta - 1)^2\frac{\langle F^2\rangle}{4\gamma T}.
\end{equation}

\noindent\textbf{Proof.} We write \(\rho(\{\xi_i\}; T)\) to denote the probability density for observing a noise history \(\{\xi_i\}\) at temperature \(T\) and \(\rho(j; T)\) the corresponding probability density for observing a current \(j\) at temperature \(T\). By contraction from \(\rho(\{\xi_i\}; T)\), we can formally write
\begin{align}
	\rho(j; T) &= \prod_i\int\text{d}\xi_i\,\rho(\{\xi_i\}; T)\,\delta(j(\{\xi_i\}) - j)\\
	    &\asymp \prod_i\int_{\Omega_j}\text{d}\xi_i\,\rho(\{\xi_i\}; T),\\
	    &\qquad \qquad \qquad \Omega_j = \{\{\xi_i\}: J(\{\xi_i\}) = j\}. \nonumber
\end{align}
The \(\delta\)-function constraint restricts the integration to the subset of trajectories which generate a current \(j\), and \(\asymp\) denotes equivalence up to sub-exponential factors (that is, ignoring the prefactor associated with transforming variables within the \(\delta\) function).
Integrating over any subset \(\Lambda_j \subseteq \Omega_j\) makes this an inequality:
\begin{equation}
	\rho(j; T) \geq \prod_i\int_{\Lambda_j}\text{d}\xi_i\,\rho(\{\xi_i\}; T).
\end{equation}
The rate function \(I(j; T)\) is a function of \(\rho(j; T)\),
\begin{equation}
    \label{eq:ratefndefn}
	I(j; T) = \lim_{\tau\to\infty}-\frac{1}{\tau}\ln\rho(j; T),
\end{equation}
so by making use of this expression, we can transform the bound into a bound on the rate function as
\begin{equation}
    \label{eq:genericratefnbound}
	I(j; T) \leq \lim_{\tau\to\infty}-\frac{1}{\tau}\ln\prod_i\int_{\Lambda_j}\text{d}\xi_i\,\rho(\{\xi_i\}; T).
\end{equation}
Asymptotically, we have
\begin{equation}
    \label{eq:probexpr}
	\rho(\{\xi_i\}; T) \asymp \exp\left(-\frac{1}{2}\sum_{i=0}^{N-1}\xi_i^2\right),
\end{equation}
and the bijection of Eq.~\eqref{eq:scalednoise} suggests that we consider the set
\begin{equation}
	\Lambda_{\langle j\rangle/\eta} = \left\{\{\xi_i\}: J(\{\xi_i\}) = \frac{\langle j\rangle}{\eta},\, \xi_i =  \tilde{\xi}_i + F_i\sqrt{\frac{\Delta t}{2\gamma T}}(\eta - 1) \right\}
\end{equation}
at temperature \(T/\eta\). Evaluating the right-hand side of the inequality gives
\begin{widetext}
\begin{align}
	\prod_i\int_{\Lambda_{\langle j\rangle/\eta}}\text{d}\xi_i\,\rho\left(\{\xi_i\}; \frac{T}{\eta}\right)&= \prod_i\int_{\Lambda_{\langle j\rangle/\eta}}\text{d}\xi_i\,\exp\left(-\frac{1}{2}\sum_{i=0}^{N-1}\left[\tilde{\xi}^2 + (\eta-1)\sqrt{\frac{\Delta t}{2\gamma T}}F_i\tilde{\xi}_i + (\eta-1)^2\frac{F_i^2 \Delta t}{2\gamma T}\right]\right)\\
		&= \prod_i\int_{\Lambda_{\langle j\rangle/\eta}}\text{d}\xi_i\,\rho\left(\{\tilde{\xi}_i\}; \frac{T}{\eta}\right)\exp\left(-\frac{1}{2}\sum_{i=0}^{N-1}\left[(\eta-1)\sqrt{\frac{\Delta t}{2\gamma T}}F_i\tilde{\xi}_i + (\eta-1)^2\frac{F_i^2 \Delta t}{2\gamma T}\right]\right)\\
		&= \left\langle\exp\left(-\frac{N}{2}\sum_{i=0}^{N-1}\left[\frac{\eta-1}{N}\sqrt{\frac{\Delta t}{2\gamma T}}F_i\tilde{\xi}_i + \frac{(\eta-1)^2}{N}\frac{F_i^2 \Delta t}{2\gamma T}\right]\right)\right\rangle\\
		&\to \exp\left(-(\eta-1)^2\frac{\tau\langle F^2\rangle}{4\gamma T}\right).
\end{align}
\end{widetext}
The last line follows in the thermodynamic (long-time) limit. In this limit, the intensive stochastic sums \(N^{-1}\sum_iF_i\tilde{\xi}_i\) and \(N^{-1}\sum_iF_i^2\) concentrate around their expectation values \(0\) and \(\langle F^2\rangle\). Plugging this result into the generic bound on \(I(j; T)\), Eq.~\eqref{eq:genericratefnbound}, gives
\begin{equation}
	I\left(\frac{\langle j\rangle}{\eta};\frac{T}{\eta}\right) \leq \frac{(\eta - 1)^2}{\eta}\frac{\langle F^2\rangle}{4\gamma T}.
\end{equation}
The factor of \(\eta\) in the denominator on the right comes from the fact that we must divide by \(\eta\tau\), rather than \(\tau\), when converting from densities to rate functions for scaled trajectories.\\

\noindent \textbf{Lemma 2.} The second derivative of the rate function with respect to \(\lambda\) along the tangent to the curve \(\langle j\rangle_T\) vanishes at the point of tangency.

\vspace{1ex}
\noindent \textbf{Proof.} The curve is parametrized as \((\lambda T_0, \langle j\rangle_{\lambda T_0})\) and its tangent line as \((\lambda T_0, \langle j\rangle_0 + \kappa(\lambda - 1)T_0)\), since by Taylor expansion
\begin{equation}
	\langle j\rangle_{\lambda T_0} = \langle j\rangle_0 + \kappa (\lambda T_0 - T_0) + O((\lambda-1)^2),
\end{equation}
where again
\begin{equation}
    \kappa := \left.\frac{\text{d}\langle j\rangle_{\lambda T_0}}{\text{d}T}\right|_{\lambda = 1}
\end{equation}
By applying the chain rule on \(I(\lambda) = I(f(\lambda), g(\lambda))\), we find that
\begin{align}
	I'(\lambda) &= I_ff' + I_gg',\\
	I''(\lambda) &= (f'\partial_f + g'\partial_g)I'\\
		&= (f')^2I_{ff} + (g')^2I_{gg} + 2f'g'I_{fg}\\
		&\qquad \qquad \qquad + f''I_f + g''I_g, \nonumber
\end{align}
where primes (\('\)) denote derivatives with respect to \(\lambda\) and subscripts denote partial derivatives with respect to the relevant parameter. We take \(f(\lambda) = \lambda T_0\), and \(g(\lambda)\) is either \(\langle j\rangle_{\lambda T_0}\) or its expansion to first order.

In this case, the two different parametrizations of \(f(\lambda)\) only differ for \(f''\) and higher derivatives of \(f\), so proving the claim is equivalent to ensuring that such terms do not appear in the expression for \(I''(\lambda)\). \(I_f \equiv 0\) because the rate function is minimized at the average value \(\langle j\rangle\), and we need not worry about \(I_{ff}\) because it depends only on the value of \(f\) rather than any of its derivatives. (The notation \(I_{ff}\) means that you differentiate I with respect to its first argument twice, then plug in \(f\).)

\(f''\) appears only in the combination \(f''I_f\), which we have argued will vanish because \(I_f \equiv 0\). Thus, it is valid to assert that \(I''(\lambda) = 0\) even if we parametrize the curve \((\lambda T_0, \langle j\rangle_{\lambda T_0})\) along its tangent line \((\lambda T_0, \langle j\rangle_0 + \kappa(\lambda - 1)T_0)\) instead.

\section{Entropy production and dynamical activity} \label{app:actionquantities}
In this appendix, we formalize the definitions of the entropy production and dynamical activity.
Consider a trajectory \(\{x_i\}\) and its time reversal \(\{\bar{x}_i\} \equiv \{x_{N-i}\}\).
For simplicity, we will consider a one-dimensional underdamped system, though these calculations generalize to multiple dimensions and the overdamped regime as well.
The probability of observing such a trajectory can be expressed in terms of its action \(\tau I(\{x_i\})\) as
\begin{equation}
	\pi(\{x_i\}) \propto \exp(-\tau I(\{x_i\})).
\end{equation}
The entropy production \(\Sigma\) and dynamical activity \(\Upsilon\) are respectively defined as the time-antisymmetric and time-symmetric components of the action.
More precisely, we have
\begin{align}
	\label{eq:actiondecomposition}
	-\tau I(\{x_i\}) &= \Upsilon(\{x_i\}) + \frac{1}{2}\Sigma(\{x_i\})\\
	-\tau I(\{\bar{x}_i\}) &= \Upsilon(\{\bar{x}_i\}) + \frac{1}{2}\Sigma(\{\bar{x}_i\})\\
	    &= \Upsilon(\{x_i\}) - \frac{1}{2}\Sigma(\{x_i\}). 
\end{align}
By combining Eq.~\eqref{eq:ratefndefn} and Eq.~\eqref{eq:probexpr}, we note that
\begin{equation}
	I(\{x_i\}) = \lim_{\tau\to\infty}\frac{1}{2\tau}\sum_{i=0}^{N-1}\xi_i^2.
\end{equation}
Furthermore, following Eqs.~\eqref{eq:contunderdampedLangevin} and \eqref{eq:contunderdampedLangevintwo},
\begin{align}
    \label{eq:compunderdampedLangevin}
    \Delta x_i &= b\Delta t\left(\dot{x}_i + \frac{F_i\Delta t}{2}\right) + \frac{b\Delta t}{2}\sqrt{2\gamma T\Delta t}\xi_i,\\
    \Delta\dot{x}_i &= \mathring{F}_i\Delta t - \gamma\Delta x_i + \sqrt{2\gamma T\Delta t}\xi_i \label{eq:compunderdampedLangevintwo}
\end{align}
and \(\mathring{F}_i \equiv (F_i + F_{i+1})/2\).
Time reversal can be effected by traversing a given trajectory backward---switching all instances of \(i\) and \(i+1\)---and changing the sign of all quantities odd in time to get
\begin{equation}
	\Delta\dot{x}_i = \mathring{F}_i\Delta t + \gamma\Delta x_i + \sqrt{2\gamma T\Delta t}\bar{\xi}_{i+1}.
\end{equation}
Hence solving for \(\Sigma(\{\bar{x}_i\})\) and \(\Upsilon(\{\bar{x}_i\})\) gives
\begin{align}
    \label{eq:entrprod}
	\Sigma(\{\bar{x}_i\}) &= \frac{1}{T}\sum_{i=0}^{N-1}\Delta x_i\left(\mathring{F}_i - \frac{\Delta\dot{x}_i}{\Delta t}\right),\\
    \label{eq:dynamicalactivity}
	\Upsilon(\{\bar{x}_i\}) &= \frac{\Delta t}{4\gamma T}\sum_{i=0}^{N-1}\left[2\mathring{F}_i\frac{\Delta\dot{x}_i}{\Delta t} - \mathring{F}_i^2 - \gamma^2\left(\frac{\Delta x_i}{\Delta t}\right)^2\right].
\end{align}
Taking the expectation values of these quantities in the steady state leads to the entropy production \(\langle\Sigma\rangle\) and dynamical activity \(\langle\Upsilon\rangle\), precise expressions for which are given in Appendix~\ref{app:expectationvalues}.
Our expression for dynamical activity differs from that in \cite{van2019uncertainty} and in the main text by the additive term \(\gamma\tau/2\) because we have employed the It\^{o} discretization for the path action rather than the Stratonovich discretization.
For ease of comparison with \cite{van2019uncertainty}, the expression for the dynamical activity in the main text is that used in \cite{van2019uncertainty}, but we derive here the expression we would otherwise have obtained for the dynamical activity.

\section{Evaluation of some expectation values} \label{app:expectationvalues}

As in Appendix~\ref{app:actionquantities}, we perform our derivations in one dimension for simplicity. For the entropy production, we would like to show that
\begin{equation}
    \langle\Sigma\rangle = \tau\frac{\gamma\langle\dot{x}^2\rangle}{T} - \gamma\tau = \tau\frac{\langle F\dot{x}\rangle}{T}.
\end{equation}
We will do this sequentially, first proving the first equality and then the second.

Taking expectations directly from Eq.~\eqref{eq:entrprod}, we have
\begin{align}
    \langle\Sigma\rangle &= \frac{1}{T}\sum_{i=0}^{N-1}\left\langle\Delta x_i\left(\mathring{F}_i - \frac{\Delta\dot{x}_i}{\Delta t}\right)\right\rangle\\
        &\approx \frac{\gamma}{T}\sum_{i=0}^{N-1}\langle\dot{x}_i^2\rangle\Delta t - \frac{1}{T}\sqrt{\frac{2\gamma T}{\Delta t}}\sum_{i=0}^{N-1}\langle\Delta x_i\xi_i\rangle\\
        &= \tau\frac{\gamma\langle\dot{x}^2\rangle}{T} - \gamma\tau + O(\Delta t).
\end{align}
In the second equality, we have simplified the expectation value using Eq.~\eqref{eq:compunderdampedLangevintwo}. In the third equality, we have simplified the first expectation value and evaluated the second by multiplying Eq.~\eqref{eq:compunderdampedLangevin} through by \(\xi_i\), then taking expectations.

It remains to prove that 
\begin{equation}
    \label{eq:tobeprovenone}
    \langle F\dot{x}\rangle = \gamma\langle\dot{x}^2\rangle - \gamma T.
\end{equation}
First, we prove the auxiliary result that
\begin{equation}
    \label{eq:auxilidentity}
    \sum_{i=0}^{N-1}\langle\dot{x}_i\Delta\dot{x}_i\rangle = -\gamma T\tau,
\end{equation}
which follows from
\begin{align}
    0 &= \sum_{i=0}^{N-1}\frac{1}{2}\langle(\dot{x}_i + \dot{x}_{i+1})\Delta\dot{x}_i\rangle\\
        &= \sum_{i=0}^{N-1}\left[\langle\dot{x}_i\Delta\dot{x}_i\rangle + \frac{1}{2}\langle(\Delta\dot{x}_i)^2\rangle\right]\\
        &= \sum_{i=0}^{N-1}\left[\langle\dot{x}_i\Delta\dot{x}_i\rangle + \gamma T\Delta t + O((\Delta t)^{3/2})\right],
\end{align}
where the first equality holds for a telescoping sum in the steady state and the final equality is obtained by direct computation from the underdamped Langevin equation, Eq.~\eqref{eq:compunderdampedLangevintwo}. Hence multiplying Eq.~\eqref{eq:compunderdampedLangevintwo} by \(\dot{x}_i\) and taking expectations gives
\begin{equation}
    \sum_{i=0}^{N-1}\langle\dot{x}_i\Delta\dot{x}_i\rangle \approx \sum_{i=0}^{N-1}\Delta t(\langle F_i\dot{x}_i\rangle - \gamma\langle\dot{x}_i^2\rangle),
\end{equation}
and combining this with Eq.~\eqref{eq:auxilidentity} gives Eq.~\eqref{eq:tobeprovenone}, which verifies the expression for \(\langle\Sigma\rangle\).

For the dynamical activity, we would like to show that
\begin{equation}
    \langle\Upsilon\rangle = \frac{\tau}{4}\left(\frac{\langle F^2\rangle}{\gamma T} - \frac{3\gamma\langle\dot{x}^2\rangle}{T} + 2\gamma\right).
\end{equation}
Taking expectations directly from Eq.~\eqref{eq:dynamicalactivity} gives
\begin{equation}
    \label{eq:dynamicalactivitycalc}
    \langle\Upsilon\rangle \approx \frac{\tau}{4\gamma T}\left[2\left\langle\mathring{F}_i\frac{\Delta\dot{x}_i}{\Delta t}\right\rangle - \langle F^2\rangle - \gamma^2\langle\dot{x}^2\rangle\right],
\end{equation}
and we have
\begin{align}
    \sum_{i=0}^{N-1}\langle\mathring{F}_i\Delta\dot{x}_i\rangle &\approx \sum_{i=0}^{N-1}\langle F_i\Delta\dot{x}_i\rangle\\
        &\approx \sum_{i=0}^{N-1}\langle F_i\mathring{F}_i\rangle\Delta t - \gamma\langle F_i\Delta x_i\rangle,
\end{align}
where the first equality follows from expanding \(F_{i+1}\) and using the fact that \(\langle\Delta x_i\Delta\dot{x}_i\rangle \sim O(\Delta t^{3/2})\) is negligible with respect to \(F_i\Delta\dot{x}_i \sim O(\Delta t^{1/2})\). The second equality follows by substituting in Eq.~\eqref{eq:compunderdampedLangevintwo}. Dividing through by \(\Delta t\) gives
\begin{equation}
    \sum_{i=0}^{N-1}\left\langle\mathring{F}_i\frac{\Delta\dot{x}_i}{\Delta t}\right\rangle \approx \sum_{i=0}^{N-1}\left[\langle F_i^2\rangle - \gamma\langle F_i\dot{x}_i\rangle\right]. 
\end{equation}
Continuing from Eq.~\eqref{eq:dynamicalactivitycalc} gives
\begin{align*}
    \langle\Upsilon\rangle &= \frac{\tau}{4\gamma T}\left[\langle F^2\rangle - 2\gamma\langle F\dot{x}\rangle - \gamma^2\langle\dot{x}^2\rangle\right],\\
        &= \frac{\tau}{4\gamma T}\left[\langle F^2\rangle - 3\gamma^2\langle\dot{x}^2\rangle + 2\gamma^2T\right],
\end{align*}
as desired. As mentioned in Appendix~\ref{app:actionquantities}, this result differs from that used in the main text by the additive term \(\gamma\tau/2\).

\bibliography{turpaperv2}

\end{document}